\documentclass[twoculom,letter]{jpsj2}

\title{Resonant Inelastic X-Ray Scattering (RIXS) in SrCuO$_{2}$}

\author{Atsushi Higashiya$^{1}$\thanks{E-mail address:higasiya@decima.mp.es.osaka-u.ac.jp}, Akihiko Shigemoto$^{1}$, Shuuichi Kasai$^{1}$, Shin Imada$^{1}$, Shigemasa Suga$^{1}$, Michael Sing$^{1}$\thanks{permanent address: Experimental physik II, Universitat Augsburg,D-86135, Augsburg, Germany}, Changyoung Kim$^{2}$, Makina Yabashi$^{3}$, Kenji Tamasaku$^{4}$ and Tetsuya Ishikawa$^{4}$}

\inst{$^{1}$Graduate School of Engineering Science, Osaka University, 1-3 Machikaneyama, Toyonaka, Osaka 560-8531\\
$^{2}$Institute of Physics and Applied Physics, Yonsei University, Seoul 120-749, Korea\\
$^{3}$SPring-8 / JASRI 1-1-1 Kouto, Mikazuki, Sayo, Hyogo 679-5198\\
$^{4}$SPring-8 / RIKEN 1-1-1 Kouto, Mikazuki, Sayo, Hyogo 679-5148
}

\abst{The resonant inelastic x-ray scattering (RIXS) was measured in an insulating SrCuO$_{2}$ with changing the incident photon energy ($h\nu$) near the Cu $1s$ absorption edge (K-edge). Complex structures and their dependences on the momentum transfer($\Delta k$) and $h\nu$ were observed. The $h\nu$ dependence of the RIXS spectra measured for a constant $\Delta k$=2.4$\pi$ has shown a remarkable enhancement near the absorption maximum and a shift of the spectral weight toward larger loss-energy with increasing $h\nu$. Also, the energy of the inelastic loss peak in the RIXS spectra for the fixed excitation at 8.993 keV, 10 eV below the absorption maximum, has shown a clear dependence on $\Delta k$. These results are in a qualitative agreement with a theoretical prediction. Full utilization of the potential of the high brilliance X ray light source at BL19LXU of SPring-8 with 27 m long insertion device will open up a breakthrough in RIXS for heavy transition metal compounds in accordance with the improvement in focussing and analyzing elements.\\}

\kword{resonant inelastic X-ray scattering, RIXS, one dimensional antiferromagnetic insulator}

\begin{document}
\maketitle
\quad Electronic structures of many transition metal compounds with strong electron correlation have intensively been studied from both theoretical as well as experimental points of view.~\cite{Taylor, Lett55} Angle-resolved photoemission spectroscopy (ARPES) has been very useful for probing occupied electronic states of metallic as well as semiconducting materials. For example a lot of high resolution studies have been performed for high $T_{\rm C}$ cuprate superconductors.~\cite{Modern} However, ARPES requires ultrahigh vacuum condition and very clean and specular single crystal sample surfaces. Still high resolution ARPES at low photon energy ($h\nu$) is rather surface sensitive and sometimes the obtained results are not consistent with the bulk electronic structures.~\cite{Suga, Sekiyama} Inverse angle resolved photoemission studies are also surface sensitive and their energy resolutions are not high enough to study  unoccupied electronic states in detail. Moreover both techniques are not applicable to insulators. Bulk intrinsic information on unoccupied electronic states together with the occupied states is required in order to fully understand the electronic structures of such materials as high $T_{\rm C}$ cuprates and colossal magnetoresistance manganites.~\cite{Science1,Science2, Lett74} For example, studies of the wave number ($k$) dependence of the Mott-Hubbard gap will provide useful information on the electron correlation effects.\\ \quad In order to satisfy both bulk sensitivity and $k$ resolution, the inelastic X-ray scattering has recently been developed.~\cite{Lett77,Lett76,Lett88} In general, the inelastic X-ray scattering from the valence charge distribution is very weak and thus difficult to be distinguished with good statistics from the total scattering signal. Therefore resonance enhancement by the excitation near the core excitation threshold is often utilized. \\ \quad We have performed the resonance inelastic X-ray scattering (RIXS) for single crystalline SrCuO$_{2}$, which is a typical one dimensional (1D) antiferromagnetic insulator. A theoretical study of RIXS for 1D copper oxides has predicted $h\nu$ and $\Delta k$ dependences of the RIXS spectra ~\cite {RevB61}. The SrCuO$_{2}$ has two Cu-O chains ( c-axis ) combined to each other by edge-sharing and is reported to show the spin-charge separation behavior with 1D nature.~\cite{Maekawa} In general, these insulating copper oxides have the lower Hubbard band (LHB) in the occupied part and the upper Hubbard band (UHB) in the unoccupied part separated by a Mott-Hubbard gap. RIXS experiment can probe the excitation across this Mott-Hubbard gap. One of the advantages of RIXS is the freedom to choose $h\nu$ near the core absorption threshold, resulting in different intermediate states. In this Letter, we report the spectral changes as functions of the $h\nu$ and momentum transfer $\Delta k$, demonstrating the resonance enhancement behavior near the Cu $1s$ edge (K-edge).\\
\begin{figure}
\begin{center}
\includegraphics[width=10cm]{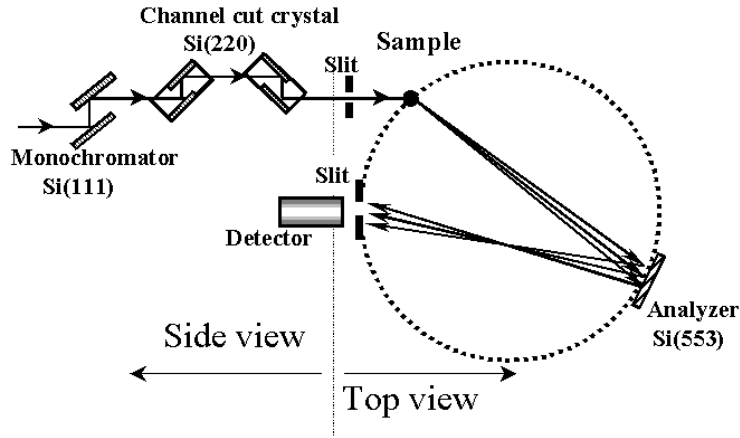}
\end{center}
\caption{Schematic illustration of the experimental setup for RIXS measurements.For simplicity, a side view and a top view are shown for the front and the rear parts, respectively.}
\label{fig1}
\end{figure}
\quad The experiment was carried out in the third hutch of the beam line BL19LXU~\cite{Lett87} at SPring-8. The linearly polarized light is delivered from an in-vacuum 27 m long insertion device. The experimental setup for RIXS is schematically shown in Fig. 1. The undulator radiation tuned to a proper $h\nu$ is monochromatized by two Si (111) crystals. It is further monochromatized by two channel-cut Si (220) crystals. The full width at half maximum (FWHM) of the energy resolution of the excitation photons is thus set to 0.3 eV. Then the monochromatic light is incident onto the polished surface of SrCuO$_{2}$ kept at room temperature in an evacuated chamber with polyimid windows. The horizontally scattered radiation by the sample with the momentum (or wave number $k$) transfer ($\Delta k$) parallel to the Cu-O chain is analyzed by a spherically bent Si (553) crystal. The diameter of the Roland circle is set to 1 m. The radiation from the analyzer is focused on the detector. The total energy resolution, determined from the quasi-elastic scattering from the sample, has a width of about 0.6 eV (FWHM). Our RIXS measurement has been performed in two modes: One for different $h\nu$ near the Cu $1s$ edge (K-edge) ( $h\nu \sim$ 9.005 keV ) with fixing $\Delta k$ to 2.4$\pi$ and the other for different $\Delta k$ with fixing $h\nu$ at 8.993 keV.\\
\begin{figure}[t]
\begin{center}
\includegraphics[width=5.5cm]{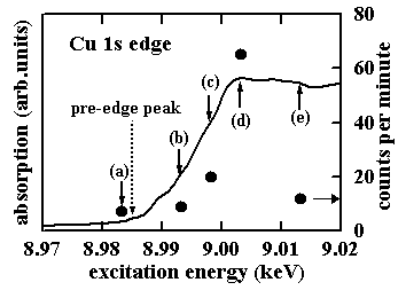}
\end{center}
\caption{Absorption spectrum of SrCuO$_{2}$. The solid arrows and labels (a)$-$(e) indicate the excitation energies at which RIXS spectra were measured.}
\label{fig2}
\end{figure}
\quad The solid curve in Fig. 2 displays the absorption spectrum of SrCuO$_{2}$ near the Cu K-edge, obtained by means of the fluorescence yield. The solid arrows in Fig.2 indicate the $h\nu$ at which the RIXS spectra are measured. The pre-edge peak indicated by the dashed arrow corresponds to the Cu $1s$-$3d$ quadrupole transition.~\cite{RevB37} In the monovalent copper compounds, the Cu $3d$ orbital is fully occupied and such an excitation is forbidden. Therefore, the observation of this pre-edge structure is consistent with the divalent character of Cu in this compound, though its intensity is not so strong. We then observe structures near 8.99 and 8.997 keV. These structures are ascribable to dipole allowed transitions from the Cu $1s$ to Cu $4p$$\pi$ states.~\cite{RevB37} The structure near the label (d) is ascribed to the Cu $1s$-Cu $4p$$\sigma$ transition.\\ \quad Figure 3 shows the inelastic scattering spectrum in a wide loss energy region for $h\nu$ = 8.993 keV ( label (b) in Fig.2 ) at $\Delta k$ = 2.2$\pi$. This measurement is performed just to know the wide energy behavior, so that no patience is paid to get a good S/N ratio. The characteristic peak near 14 eV is identified as the resonant Raman scattering related to the excitation from the Cu $3d$ to $4p$ states or to the Cu K$\beta$$_{5}$ emission line.~\cite{Lett80}\\
\begin{figure}[t]
\begin{center}
\includegraphics[width=5.5cm]{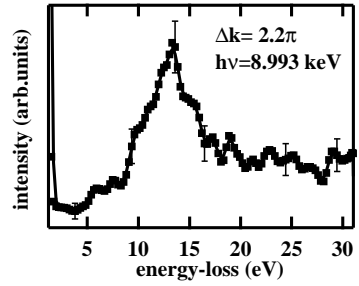}
\end{center}
\caption{The wide energy-loss spectra observed at $h\nu$= 8.993 keV. The momentum transfer $\Delta k$ is 2.2$\pi$.}
\label{fig3}
\end{figure}
\begin{figure}[b]
\begin{center}
\includegraphics[width=4cm]{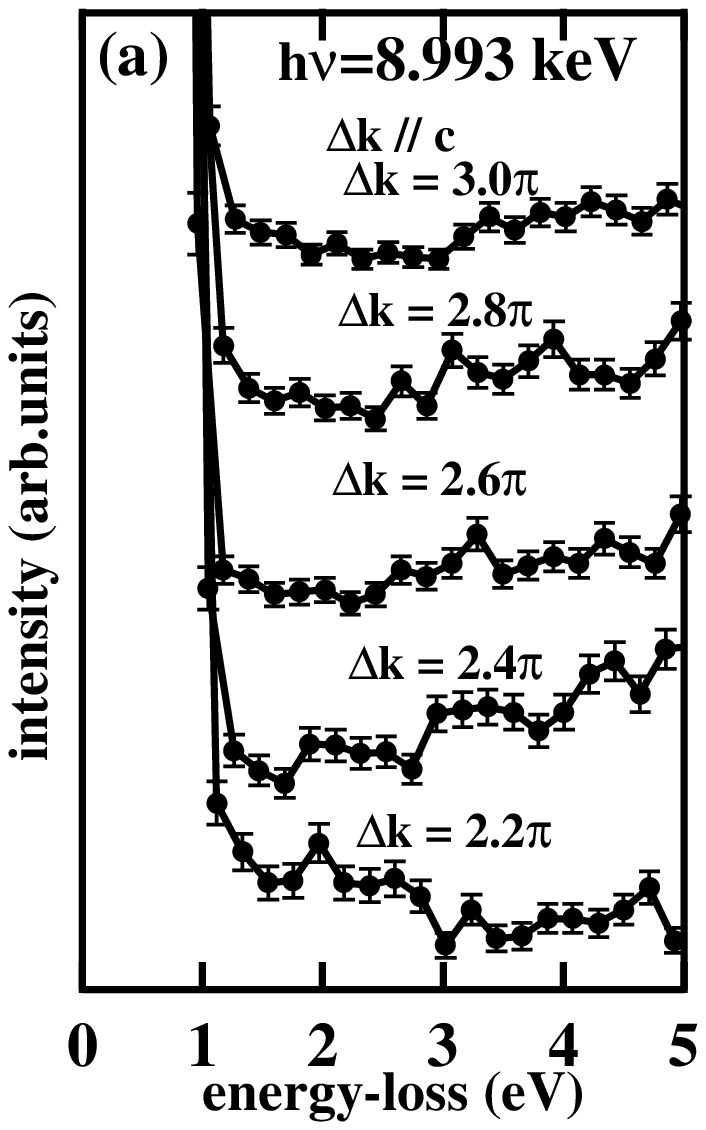}
\includegraphics[width=4cm]{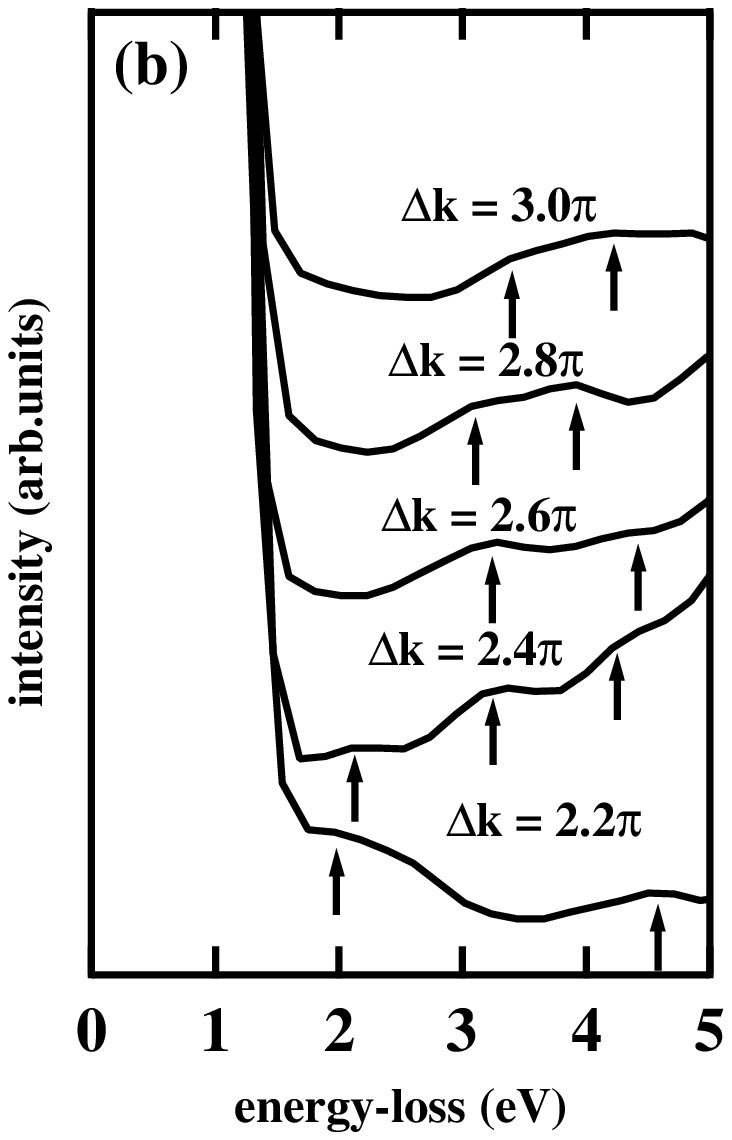}
\includegraphics[width=4cm]{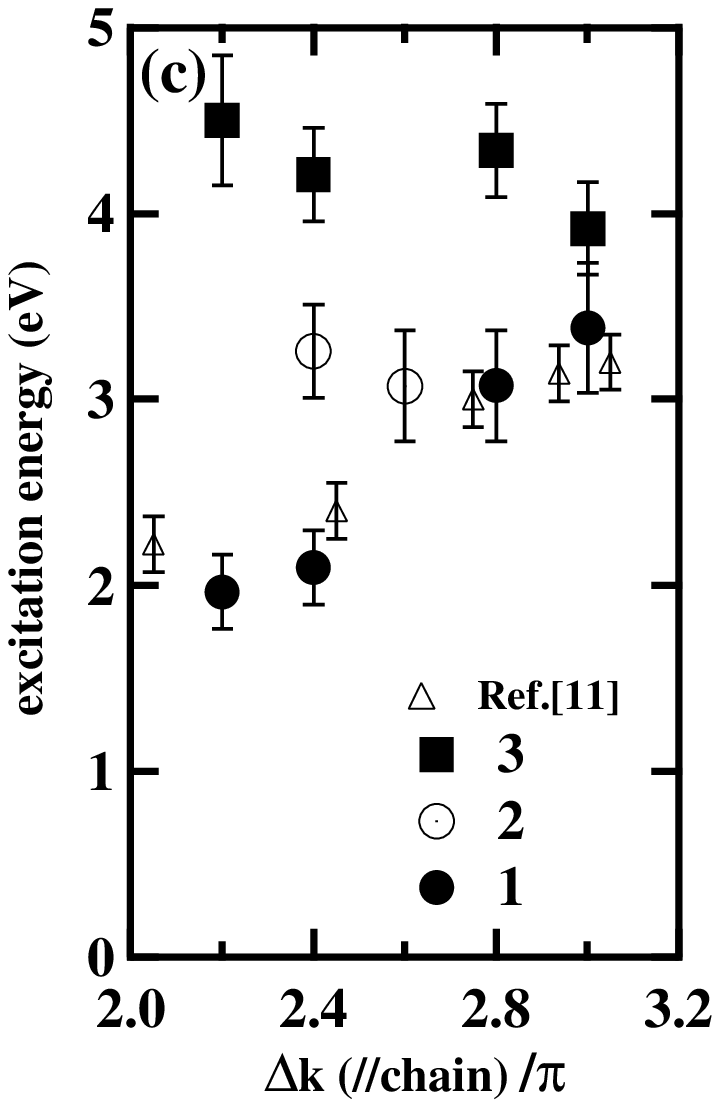}
\end{center}
\caption{(a)Dependence of the RIXS spectra on the momentum transfer $\Delta k$ along the chain direction for a fixed $h\nu$ = 8.993 keV. The (b)is the fir results. (c)Plot of the peak positions against $\Delta k$. The results by Hassan et al.~\cite{Lett88} are included for comparison.\\}
\label{fig4b}
\end{figure}
 \quad The $\Delta k$ dependence of the RIXS spectra was measured at $h\nu$ = 8.993 keV as summarized in Fig. 4. The raw data spectral shapes clearly show the $\Delta k$ dependence. However, the S/N ration is not sufficient to determine the peak energies. This is because the employed $h\nu$ is much below the absorption peak energy and therefore the resonance enhancement is not so strong (see Fig. 2). Therefore, we employed the least square fit by assuming Gaussian peaks with the FWHM = 0.6eV, which could simulate the elastic peak. The fit results are shown in Fig.4(b). Still multiple peaks are observed for each $\Delta k$. The spectrum at $\Delta k$=2.2$\pi$ has a structure overlapping with the tail of the elastic peak. So the energy of this peak is determined from the second energy derivative. The energy positions of the peak indicated by the arrows are plotted in Fig.4(c) together with the result of ref.11. The error bars were determined by FWHM of each peak. There seem to be three dispersions represented by the marks 1, 2 and 3. The peak 1 shifts toward larger loss energy with the increase of $\Delta k$. The peak 2 separated from the peak 1 for $\Delta k$=2.4$\pi$ merges with the peak 1 for $\Delta k$=2.6$\pi$ and 2.8$\pi$. The peak 3 does not shift much with $\Delta k$. These whole results are in qualitative agreement with Fig.1 of ref.11. Though the statistics of our present results are slightly worse than those in ref.11, the minimum energy in our experiment is slightly lower than that of the previous results.\\
 \begin{figure}
\begin{center}
\includegraphics[width=4.2cm]{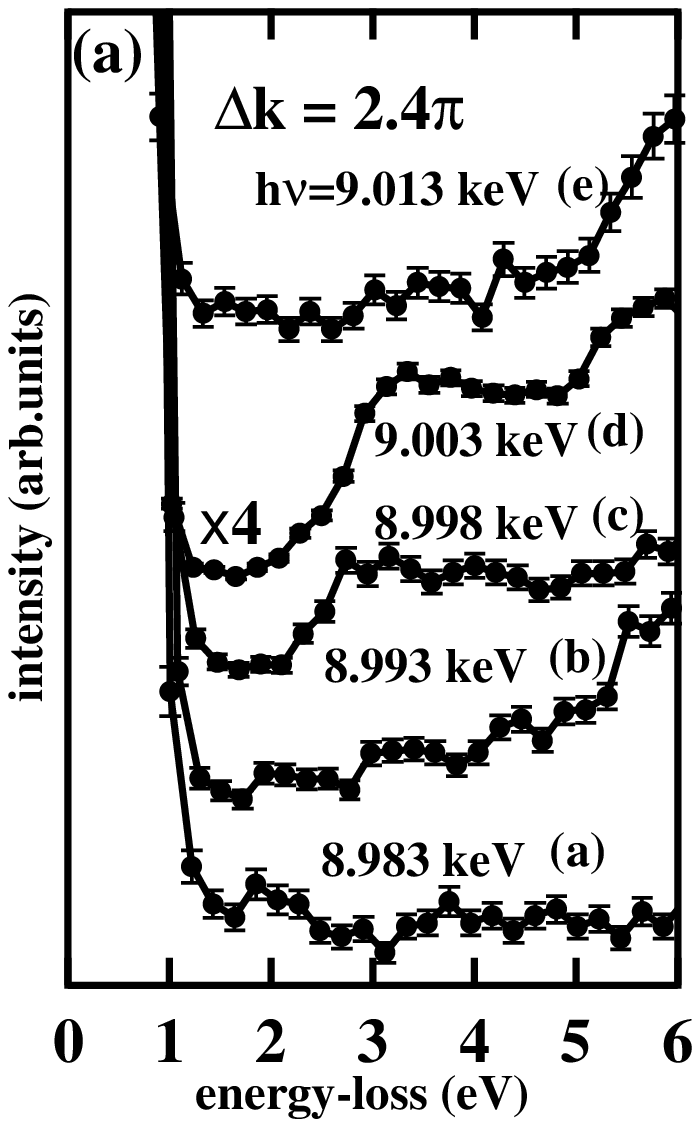}
\includegraphics[width=3.8cm]{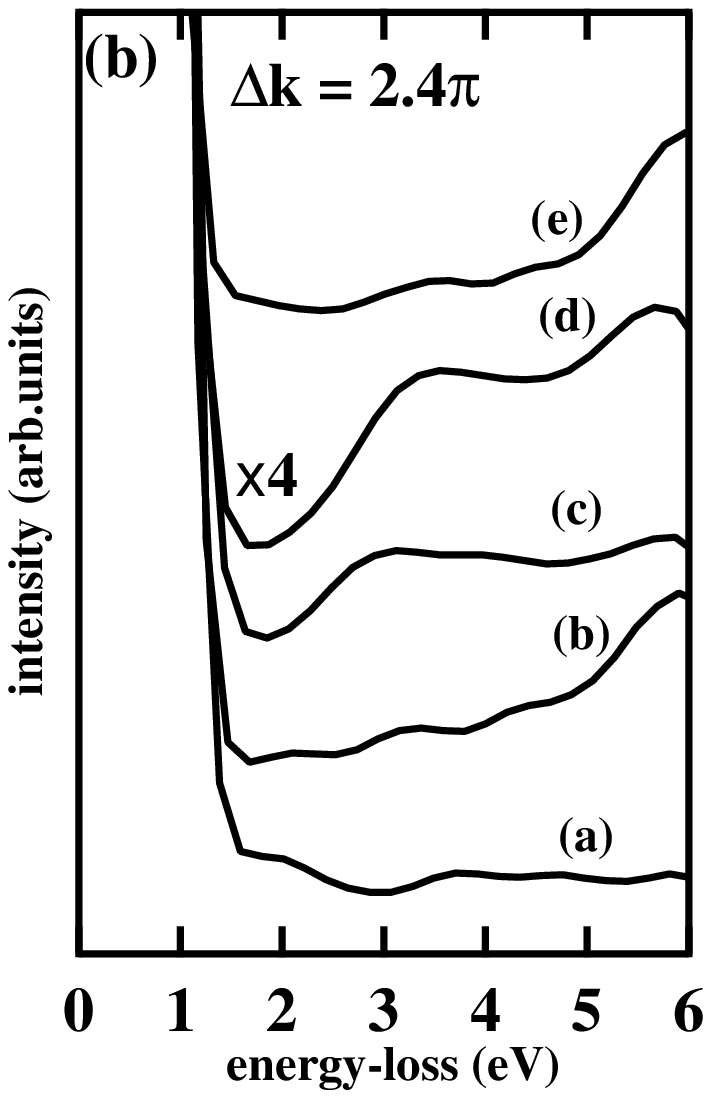}
\end{center}
\caption{$h\nu$ dependence of the RIXS spectra of SrCuO$_{2}$ for a fixed momentum transfer of 2.4$\pi$. The spectrum (d) is shown after the reduction of a factor 4.\\}
\label{fig5}
\end{figure}
\quad Then the $h\nu$ dependence of RIXS was measured for fixed $\Delta k$ = 2.4$\pi$ as summarized in Fig. 5. The raw data spectral shapes clearly show the $h\nu$ dependence. We again employed the least square fit by assuming Gaussian peaks. The fit results are shown in Fig.5 (b). In Figs. 5 (a) and (b), the spectrum (d) is shown after the reduction of a factor 4. This means that the inelastic peak intensity is greatly enhanced at the excitation very near the absorption maximum. On the other hand, the spectrum (a) shows a small structure at even larger loss energy than the result at $h\nu$ = 8.993 keV and $\Delta k$ = 2.2$\pi$. The spectrum (b) has three structures in the region within 5 eV and has a large structure near 6 eV. The spectra (c) and (d) have a large structure in the region between 2.5 and 4.5 eV. The excitation energy for (e) is far away from that of (d) and the loss peak within 5 eV becomes very weak again compared with the results for the excitations (c) and (d). The spectra obtained under the excitations (a), (b), (c), and (e) show only 5 or 6 counts per channel and per minute for the loss region within 5 eV. On the other hand, the count of the spectrum obtained under the excitation (d) has more than 60 counts per minute (right scale in Fig. 2.). The $h\nu$ dependence of the intensity distribution of the RIXS spectra is already predicted in ref.11. The present experimental results for  $\Delta k$ = 2.4$\pi$(0.4$\pi$ offset from 2$\pi$) with the shift of the spectral weight toward larger loss energy with increasing $h\nu$ is consistent with this theoretical prediction.\\  
\quad Besides, the structure near 6 eV may be ascribed to a charge transfer excitation from the ground state to the antibonding-type excited state.~\cite{Lett88} It is already mentioned that the spectral shape within 5 eV changes much with $h\nu$. The loss energy peak near 2 eV is observed for (a) and (b), whereas it becomes less obvious for the excitations (c), (d), and (e). In the case of (a), the excitation has a strong character of the quadrupole Cu $1s$-Cu $3d$ excitation and the observed structure near 2 eV can be assigned to the excitation from LHB to UHB. The loss peak near 2 eV in (b) may reflect either the contribution of the quadrupole Cu $1s$-Cu $3d$ excitation or the interatomic hybridization between the Cu $3d$ and the Cu $4p$$\pi$ orbitals. For $h\nu$ in (c) and (d), the contribution of the quadrupole excitation is negligible and the RIXS are strongly influenced by the Cu $1s$-Cu $4p$$\sigma$ dipole excitation. Then the structure near 2 eV is much suppressed due to the different symmetry of the intermediate state. The Mott gap excitation below 2eV is consistent with the bulk sensitive ARPES with the strong peak near 0.95eV at $\pi$/2 corresponding to the LHB.~\cite{suga2}\\
\quad In order to do a full-dress study of RIXS processes in SrCuO$_{2}$, we plan to do $\Delta k$ resolved measurements at several different $h\nu$ with much higher energy resolution by improving the analyzer crystals and fully utilizing the high intensity of the powerful 27 m long insertion light source.\\ 
\quad In this Letter, we have reported on RIXS in insulating SrCuO$_{2}$. It is found that RIXS spectra change much with $h\nu$ reflecting the character of the intermediate states. The $\Delta k$ resolved RIXS results are consistent with the bulk sensitive soft X-ray ARPES as well as a theoretical prediction based on an extended Hubbard model.\\ 
\quad We thank Mr. D. Miwa for his technical support. Discussions with Prof.S.Maekawa, Prof.T.Tohyama and Dr.K.Tsutsui are deeply acknowledged. This experiment was performed at SPring-8 under the proposal of 2003A0598-ND3-np. The authors are much obliged to MEXT, Japan for the support of Grant-in-Aid for COE Research.\\

\end{document}